# Restoring speech intelligibility for hearing aid users with deep learning

Peter Udo Diehl[1,2,*], Yosef Singer[1], Hannes Zilly[1], Uwe Schönfeld[2], Paul Meyer-Rachner[1], Mark Berry[1], Henning Sprekeler [3,4,5], Elias Sprengel[1], Annett Pudszuhn[2], Veit M. Hofmann[2]

* Corresponding author
[1] Audatic, Berlin, Friedrichstr. 210, 10117 Berlin, Germany
[2] Charité – Universitätsmedizin Berlin, corporate member of Freie Universität Berlin, Humboldt-Universität zu Berlin, and Berlin Institute of Health, Department of Otorhinolaryngology, Head and Neck Surgery, Campus Benjamin Franklin, Germany
[3] Department for Electrical Engineering and Computer Science, Technische Universität Berlin, Berlin, Germany
[4] Bernstein Center for Computational Neuroscience Berlin, Philippstr. 13, 10115 Berlin, Germany
[5] Exzellenzcluster Science of Intelligence, Technische Universität Berlin, Marchstr. 23, 10587 Berlin, Germany

## Abstract
**Almost half a billion people world-wide suffer from disabling hearing loss. While hearing aids can partially compensate for this, a large proportion of users struggle to understand speech in situations with background noise. Here, we present a deep learning-based algorithm that selectively suppresses noise while maintaining speech signals. The algorithm restores speech intelligibility for hearing aid users to the level of control subjects with normal hearing. It consists of a deep network that is trained on a large custom database of noisy speech signals and is further optimized by a neural architecture search, using a novel deep learning-based metric for speech intelligibility. The network achieves state-of-the-art denoising on a range of human-graded assessments, generalizes across different noise categories and – in contrast to classic beamforming approaches – operates on a single microphone. The system runs in real time on a laptop, suggesting that large-scale deployment on hearing aid chips could be achieved within a few years. Deep learning-based denoising therefore holds the potential to improve the quality of life of millions of hearing impaired people soon**.

## Introduction

Hearing loss is a debilitating condition that is associated with a large range of negative health outcomes[3], including higher levels of social isolation, dementia, depression, cortical thinning[4] and increased mortality[5]. Nevertheless, over 80% of people who would benefit from use of hearing aids do not wear them[6], with the majority of hearing aid owners who do not wear them citing difficulties with hearing in noisy situations as a main problem[1,2].

Noise reduction in most commercially available hearing aids is done by introducing spatial selectivity, e.g., by beamforming, which improves speech intelligibility for frontal sources in the presence of predominantly non-frontal noise[7–10]. Non-spatial (single microphone) noise reduction algorithms employed in hearing aids have so far not been able to provide improvements in speech intelligibility[7,11–16]. A few recent studies have shown that deep learning-based denoising[17–19] or separation of multiple competing speakers[20,21] can provide improvements in speech intelligibility for cochlear implant users[22] and hearing aid users with severe-profound[6] hearing loss under fixed signal-to-noise (SNR) conditions[17–19]. For the majority of hearing aid users, with less severe hearing loss[6], it is more challenging to provide intelligibility improvements through denoising. A very recent study has shown the ability of a deep learning based denoising system to moderately improve speech intelligibility for hearing aid users[23]. Our work builds upon these exciting results and demonstrates that deep learning based denoising may be used to provide large improvements in speech intelligibility for hearing aid users in the near future.

In order to be adopted into real-world use in hearing aids, the denoising system needs to work for: 1) a wide variety of a-priori unknown speakers, noise types, and SNR values; 2) real-time processing; 3) users with mild to severe hearing loss. In this study we present a new deep learning-based denoising system that simultaneously addresses all of those points. Based on 150,000 human ratings of three datasets covering a wide range of speakers, noises, and SNRs, our model improves upon state-of-the-art denoising systems, while being able to run on a laptop in real-time. Most importantly, in live intelligibility tests with dynamically adapting SNRs, our system improves speech intelligibility for hearing aid users with moderate-severe hearing loss to levels comparable to those for normal hearing listeners without our denoising system.

## Results

**Deep learning-based noise reduction**

Our noise reduction system comprises three key components: a denoising network; metrics that reflect human auditory perception; and an algorithm to find network architectures that maximally improve the quality of noisy speech as determined by these metrics (Figure 1). The denoising network has a U-Net[24] architecture, and predicts a complex-valued ideal ratio mask from the short-time Fourier transform of noisy speech signals (see Methods). The U-Net is trained on tens of thousands of hours of noisy speech to enhance the speech signal and mask unwanted background noise using a mean-squared error loss. Since this loss does not reflect human perception well, we evaluate the network performance using a novel deep-learning based metric that estimates human acoustic perception. To optimize the U-Net architecture we performed an evolutionary architecture search[25–28] guided by our deep-learning metric.

Human auditory perception of the quality of algorithmically synthesized content is commonly assessed by "mean opinion scores" (MOS) from human users. As obtaining MOS is time-consuming and expensive, they are typically only used to compare the performance of algorithms once they have been optimized, but not for the optimization procedure itself[29,30]. In particular, human MOS are prohibitively expensive to guide an architecture search, where hundreds or thousands of prototype networks must be evaluated. We address this challenge by gathering approximately 100,000 MOS from human subjects and training a neural network to predict the MOS given a noisy or a denoised speech sample and the associated clean version (Methods). The predicted MOS (pMOS) is then used to guide the evolutionary search algorithm (Figure 1), since it approximates human perception well and can be computed quickly enough to act as a target in the optimization loop of the neural architecture search (Figure 1). We ran the neural architecture search until improvement in performance plateaued, after which we trained the best performing network over 2.2 million steps (the equivalent of over 2 years of audio).

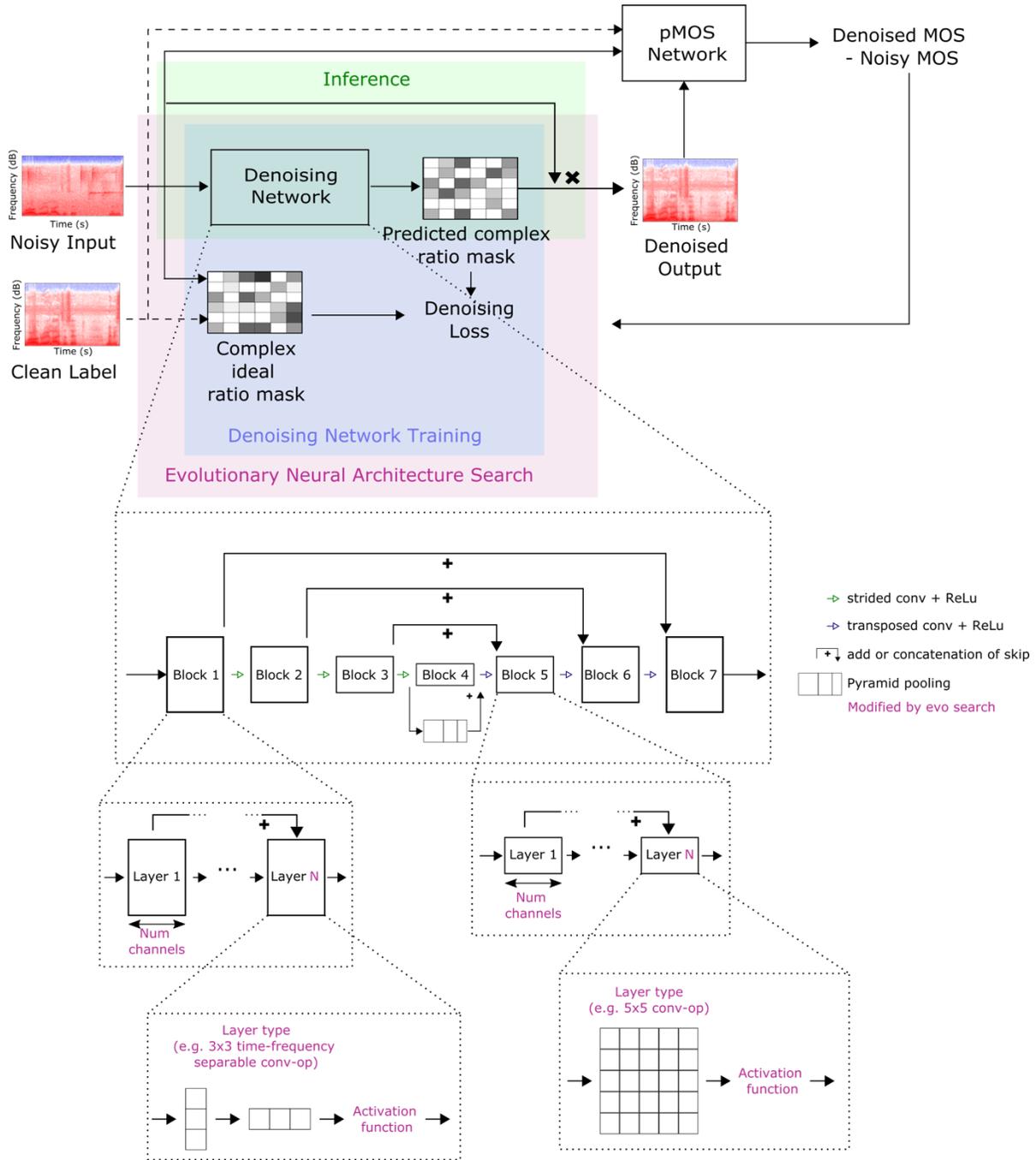

Figure 1. **Training pipeline of the denoising system using a mean opinion score (MOS)-estimator-guided neural architecture search.** The denoising network is trained to predict denoised outputs from mixed speech and noise input STFTs. To optimize the remaining error for human acoustic perception, the denoising network architecture and hyperparameters are selected by an evolutionary neural architecture search[25–28]. This search is guided by an MOS estimator, which is a deep neural network trained on a dataset generated from around 100,000 human rated audio files.

**Human evaluation of „deep" denoising**

We compared our denoising algorithm to state-of-the-art deep learning-based models (Sound of Silence, Demucs, and MHANet)[31–33] using three test datasets, one of which (*Valentini*[34]) is commonly used as a performance baseline. We created two other test datasets (*WHAMVox easy* and *WHAMVox hard*) from the publicly available VOXCeleb2[35] speech and WHAM![36] noise datasets (see Methods) to provide a more varied set of speakers, noise types and a wider range of SNR values. The test sets[34*] and example processed sound files[†] are available online.

We collected MOS ratings from human listeners for the same sound samples with noisy speech (as a baseline), enhanced by our denoising algorithm, and enhanced by the comparison models (Figure 2A). Our denoising system achieves higher human MOS scores than the comparison models on all three datasets ($p < 0.0005$, Wilcoxon signed-rank test).

Human listeners rated the samples processed with our denoising model as better than other models for a wide range of signal-to-noise ratios (SNRs; Figure 2B), which is crucial for translating the models to real-world applications. In particular, we provide substantial improvements for the SNR range between -5 and 0 dB, where the other models achieve limited improvements. Our algorithm restores the quality of sounds with -5 dB SNR to the level of unprocessed samples at 7 dB SNR, reflecting an important use-case for hearing aid users in a very busy bar or café. Processed sounds at 1 dB SNR show the same human rated MOS as clean unprocessed sounds at 20 dB SNR or more. Finally, our denoising system also improves the quality of almost clean sounds (Figure 2B at 20 dB SNR), e.g., by reducing breathing sounds and the often present white-noise during a recording.

We also compare the models on commonly used computational metrics and our pMOS estimator. The improvements achieved in human MOS scores (Figure 2A, B) were not consistently reflected in similar increases in the computational metrics (Figure 2C). In some cases, computational metrics conflicted with human perception (e.g., Valentini dataset, CSIG/COVL vs. human MOS). Nevertheless, pMOS accounts better for human perception than traditional speech quality metrics (correlation 0.9 between human MOS and pMOS vs 0.82-0.86 for PESQ, CSIG, COVL, CBAK; see Methods). These differences highlight the need to validate denoising methods on human data to avoid potential overfitting to specific computational metrics.

---

[*] The datasets can be downloaded from: https://audatic-team.github.io/WHAMVox/
[†] Sound files are available at: https://jelly-crush-c64.notion.site/Restoring-speech-intelligibility-for-hearing-aid-users-with-deep-learning-495b365b86bb406694b6efc926c78178

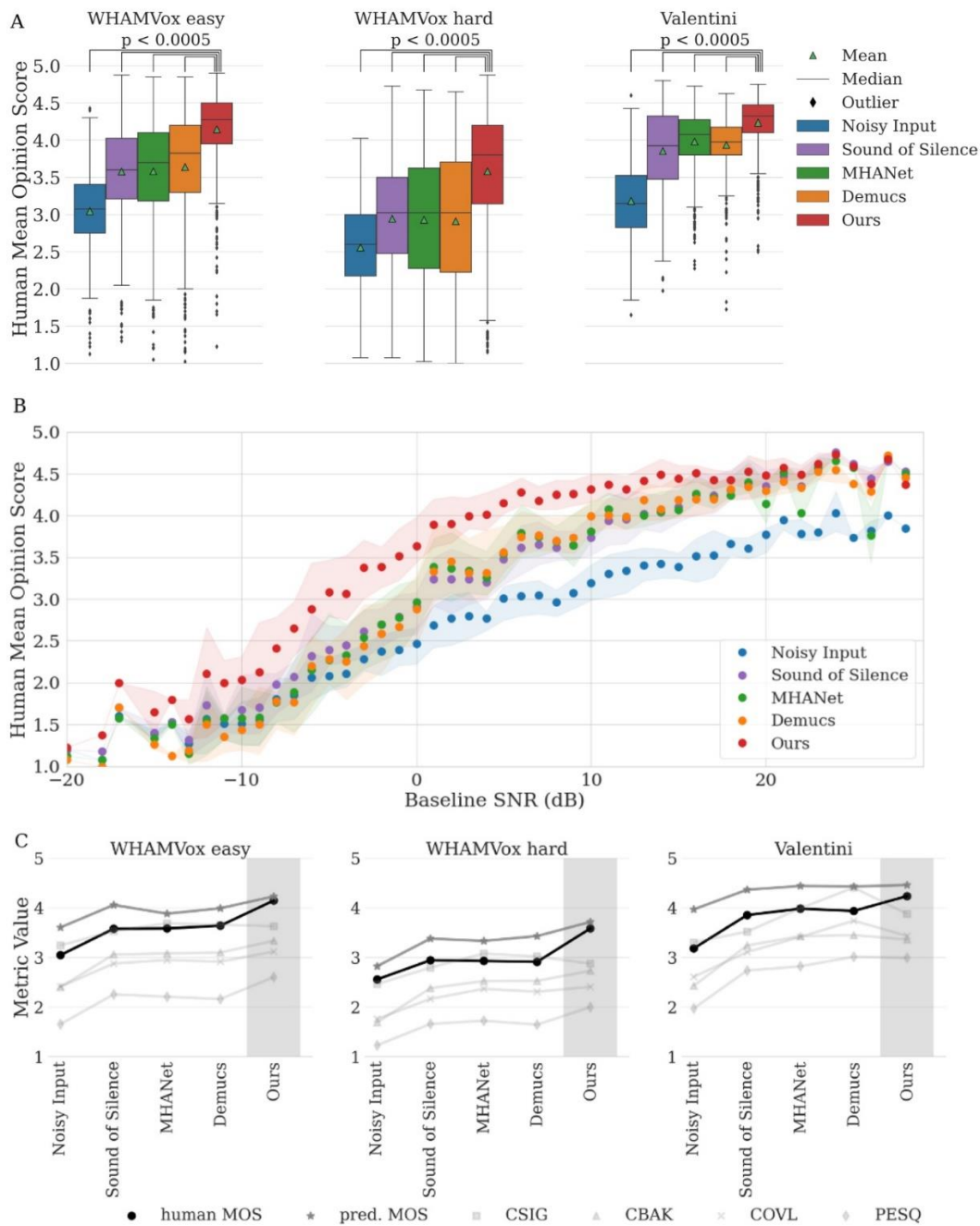

Figure 2. **Human mean opinion scores (MOS) of state-of-the-art denoising methods.** A) Comparison between current denoising models on 3 publicly available test sets[34], total of 150,000 Human MOS (500 sound files per dataset, 20 human raters per file, 5 models). B) Dependence of Human MOS on signal-to-noise ratio (1500 samples of Figure 2A). Shading: 25th and 75th quartiles, no shading for SNR values with too few audio samples. SNR values rounded to the nearest integer value. C) Comparison of denoising methods using other common speech quality metrics on a 1 to 5 scale.

**Improving speech intelligibility**

Current noise suppression systems commonly used in hearing aids improve hearing comfort, but do not improve speech intelligibility[7,11–16]. To assess the impact of our denoising system on speech intelligibility, we used the Oldenburger Satz (OLSA) test[37] (see Methods). The OLSA test measures individual speech reception thresholds (SRT), defined as the SNR at which a subject correctly identifies 50% of words. Lower SRT is better. We tested three different noise conditions, mixing clean speech with: 1) speech-shaped 'OLSA' noise, 2) restaurant noise, and 3) traffic noise. For each mixed sample we tested with and without applying our denoising system, and for normal and hearing impaired listeners, i.e., a total of 12 conditions (Figure 3A). Hearing impaired subjects wore their own hearing aids, adjusted to their individual hearing profile.

Without noise suppression, normal hearing subjects have a median SRT of -5.8 dB for OLSA noise (mean -5.6 dB, median absolute deviation 0.48). The SRTs of hearing impaired subjects are higher with a median of -3.3 dB (mean -2.1 dB, median absolute deviation 2.4). Activating our denoising system provides median changes of -3.5 dB, -3.5 dB, and -2.8 dB for hearing impaired subjects (mean changes of -4.0 dB, -4.2 dB, and -4.3 dB), respectively for the OLSA, restaurant and traffic noise ($p<0.0005$, Figure 3A, B). These improvements are greater than those reported for any other single-channel noise reduction system. For OLSA noise, they bring the median SRTs of hearing impaired subjects to –6 dB SNR, i.e. levels that are comparable to those normal hearing listeners (Figure 3A, left). In fact, for all three noise types, SRTs are not significantly different between normal hearing subjects without noise suppression and hearing aid users with denoising ($p>=0.1$; Figure 3A). Moreover, even normal hearing users' median SRTs changed by -1.9 dB, -2.5 dB, -1.3 dB (mean of -2.0 dB, -2.2 dB, and -1.5 dB SNR), for the three noise types when denoising was applied ($p<0.0005$; Figure 3A, B). The amount of improvement is consistent across the three noise types (Figure 3B). Despite large differences in SRTs across noise types before denoising, there are no significant differences in the amount of improvement across noise types within the normal and hearing impaired groups (Friedmann test: $p>0.05$ for both).

Listeners with more difficulty in noisy situations receive greater benefit from the denoising system (Figure 3C), as indicated by the inverse correlation between the improvement of the SRT and the SRT without the denoising system for all subjects. Looking at the results for individual listeners, the smallest improvement for the hearing impaired population is -0.9 dB (OLSA noise) for a subject with a middle ear disorder who has a near-normal SRT of -5.1 dB without our denoising system. Conversely, we measured the strongest improvement of -16.8 dB (traffic noise) for the subject with the strongest hearing loss, who had a SRT above 10 dB without our denoising system. This suggests that with denoising, speech understanding becomes more consistent across subjects with different degrees of hearing loss. Indeed, the median absolute SRT deviation, i.e. the variation of speech thresholds, is reduced from 2.4 dB without denoising to 1.2 dB with denoising for OLSA noise. Thus, our denoising system reduces (negative) outliers, thereby leading to an improved "worst-case" outcome for normal and hearing impaired listeners.

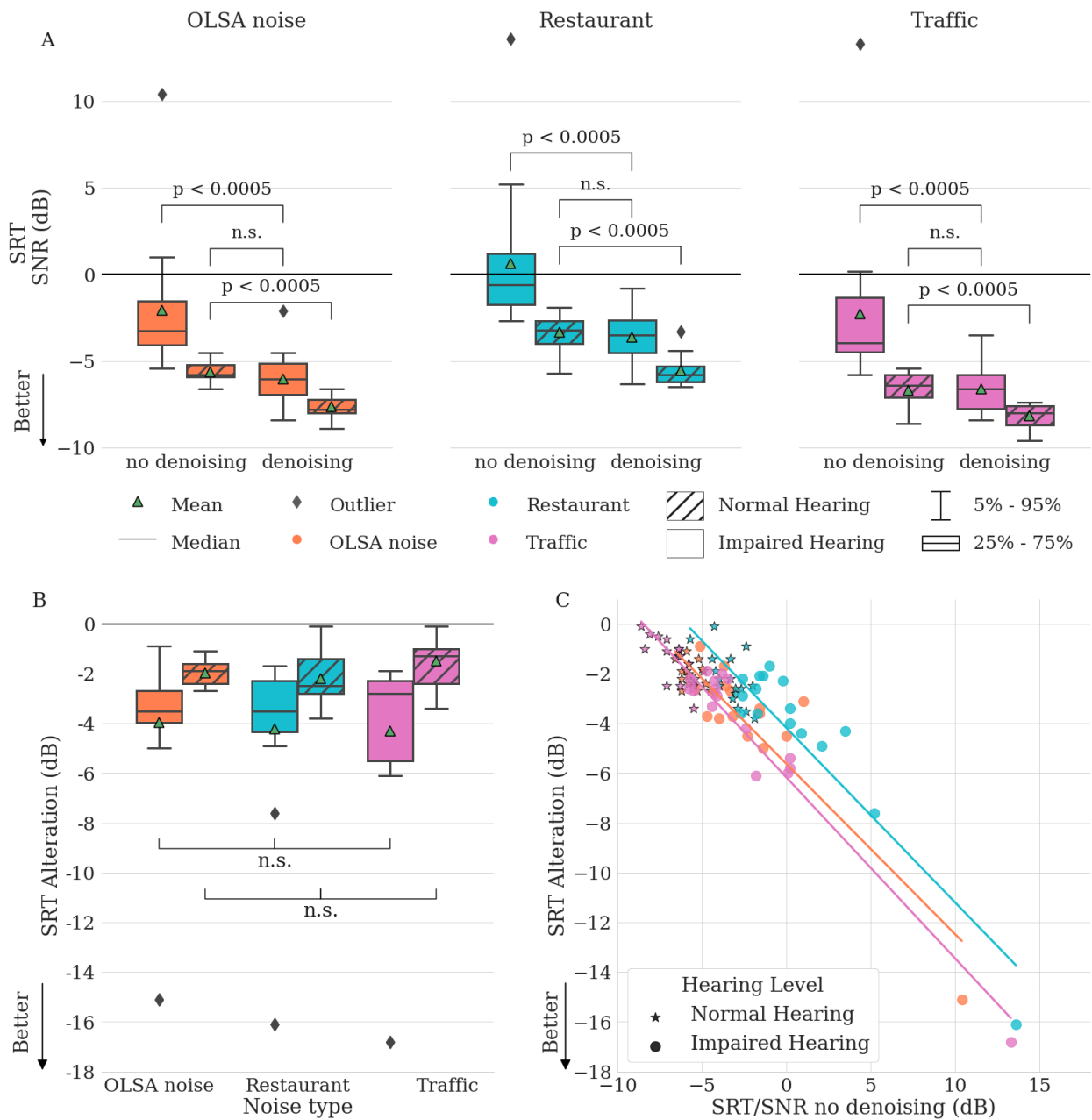

Figure 1. **Reducing speech reception thresholds (SRT) on the OLSA test using our denoising system** for hearing impaired (n=16) and normal hearing (n=17) subjects. A) SRTs for different noise types. Significance levels are shown for Wilcoxon signed-rank test for matched pairs and Mann-Whitney U test for independent samples. B) SRT improvements for each of the six testing conditions (Friedman test). C) SRT improvement vs SRT without denoising. Linear regression fits for each of the three noise type measurements, across all subjects (slopes m=-0.69, m=-0.70, m=-0.73).

# Discussion

In summary, we have shown that the suggested noise suppression system improves speech intelligibility for hearing impaired subjects to levels similar to normal hearing subjects, across a wide range of noise conditions. Almost all hearing impaired subjects improved to a similar level of speech understanding, indicating that the benefits afforded by our system are larger for individuals with a higher need.

Speech intelligibility could be further enhanced by combining single-channel denoising algorithms such as ours with existing multi-channel solutions such as beamforming. Moreover, we expect that further improvements could be achieved by adapting the sound quality metric to the specific perceptual demands of hearing aid users by gathering MOS data from hearing impaired listeners, whose perception can be quite different from normal subjects. This would allow networks to be fine-tuned to hearing aid or cochlear implant users, opening possibilities to further improve speech intelligibility in noisy environments to beyond what is currently possible with hearing aids. Additionally, our denoising system currently has no in-built system that could identify the target speaker from a mixture of multiple speakers. Thus, our system treats multiple speakers simply as "speech" and does not attenuate any of them if they are all prominent, but suppresses simultaneous background noise. For example, if two competing talkers partially overlap with a similar loudness in a café with ongoing background babbling, the two talkers would be audible but not the babbling. Separating multiple competing speakers is an active area of research[38–40] and forms an additional component of the system that could be added in future iterations. For practical use, the output of our denoising system would be mixed with the input signal, e.g. 90% denoised and 10% original, to avoid a feeling of isolation of the user from its environment and to provide a noise "reduction" instead of noise removal. This parameter could also be easily changed according to user preferences.

The improvement of speech intelligibility that can be obtained by denoising system like ours will naturally be limited by the ability of the user to understand speech in quiet environments. Similarly, our system often reaches its performance limit for very low SNRs, in which even normal hearing listeners struggle to understand speech. Conversely, as long as normal hearing subjects are able to discern the speech signals, our denoising system is able to reduce noise while maintaining high speech quality. This is a possible explanation for why our denoising system consistently improves speech intelligibility for hearing impaired listeners in noisy situations up to levels comparable to those for normal hearing listeners without the system, but barely beyond that.

Studies involving traditional noise reduction algorithms available in current hearing aids have commonly failed to show improvements in speech intelligibility for hearing aid users without relying on spatial information[7,11–16]. This is consistent across experimental conditions, including hearing aid type, speech and noise stimuli, noise reduction algorithm, language, and testing paradigm[7,11–16]. Despite this, such algorithms are commonly used in hearing aids because they offer an improvement in comfort and ease of listening[7,13,14], and a decrease in the cognitive load required to concentrate on speech in noisy environments[41–43]. While we do not test for it here, we also expect improvements in cognitive load with our system, considering that it increases the effective SNR of noisy sounds by up to 16dB[43]. In contrast to studies using traditional denoising algorithms, a recent study has shown that deep learning based denoising[23] on hearing aids can improve speech intelligibility for hearing impaired subjects. Our work is in line with other research in this field[17–19] and shows the improvements to speech intelligibility that can be made with more powerful neural network models.

## Outlook

A challenge for neural network-based denoising algorithms is their computational cost compared to denoising methods traditionally used in hearing aids. As with most deep learning-based systems, the performance of our network improves with the available computational resources. Here, we limited the size of the network such that the algorithm runs in real time on a laptop. Hence, the computational power required to achieve the presented results is higher than what is available in current hearing aids and scaling the technology to the point where it can fit in a hearing aid still requires engineering more powerful hardware and/or more efficient models. However, the gap is not prohibitively large, and we speculate that Moore's law and the exponential improvement in computational power per watt[44] should lead to a feasible implementation on a hearing aid within a few years. Additionally, the rapid progress in the algorithmic efficiency of neural networks[45,46] should further shorten adoption time.

In summary, we have presented a denoising system that enables hearing aid users to achieve speech-in-noise intelligibility levels comparable to those for normal hearing listeners and generalizes across noise environments. Deep learning-based denoising systems could hence facilitate an entirely new type of hearing improvement that is directionally independent and could prove useful not only for hearing impaired users, but also for normal hearing listeners who wish to reduce noise in noisy situations, such as crowded restaurants or bars.

## References

1. Hougaard, S. & Ruf, S. EuroTrak I: A Consumer Survey About Hearing Aids in Germany, France, and the UK. *Hearing Review* 9 (2011).

2. Hartley, D., Rochtchina, E., Newall, P., Golding, M. & Mitchell, P. Use of hearing AIDS and assistive listening devices in an older Australian population. *Journal of the American Academy of Audiology* **21**, 642–653 (2010).

3. Cunningham, L. L. & Tucci, D. L. Hearing Loss in Adults. *The New England journal of medicine* **377**, 2465–2473 (2017).

4. Ha, J. *et al.* Hearing loss is associated with cortical thinning in cognitively normal older adults. *European Journal of Neurology* **27**, 1003–1009 (2020).

5. Fisher, D. *et al.* Impairments in hearing and vision impact on mortality in older people: the AGES-Reykjavik Study. *Age and Ageing* **43**, 69–76 (2014).

6. World Health Organization. *World report on hearing*. (World Health Organization, 2021).

7. Boymans, M. & Dreschler, W. A. Field trials using a digital hearing aid with active noise reduction and dual-microphone directionality. *Audiology* **39**, 260–268 (2000).

8. Picou, E. M., Aspell, E. & Ricketts, T. A. Potential Benefits and Limitations of Three Types of Directional Processing in Hearing Aids. *Ear and Hearing* **35**, 339–352 (2014).

9. Appleton, Jennifor & König, Gabriel. Improvement in Speech Intelligibility and Subjective Benefit with Binaural Beamformer Technology. *Hearing Review* (2014).

10. Froehlich, N Matthias, Freels, Katja, & Powers Thomas A. Speech Recognition Benefit Obtained from Binaural Beamforming Hearing Aids: Comparison to Omnidirectional and Individuals. *Audiology Online* (2015).

11. Chong, F. Y. & Jenstad, L. M. A critical review of hearing-aid single-microphone noise-reduction studies in adults and children. *Disability and Rehabilitation: Assistive Technology* **13**, 600–608 (2018).

12. Völker, C., Warzybok, A. & Ernst, S. M. A. Comparing Binaural Pre-processing Strategies III: Speech Intelligibility of Normal-Hearing and Hearing-Impaired Listeners. *Trends in Hearing* (2015).

13. Brons, I., Houben, R. & Dreschler, W. A. Effects of noise reduction on speech intelligibility, perceived listening effort, and personal preference in hearing-impaired listeners. *Trends in Hearing* **18**, (2014).

14. Zakis, J. A., Hau, J. & Blamey, P. J. Environmental noise reduction configuration: Effects on preferences, satisfaction, and speech understanding. *International Journal of Audiology* **48**, 853–867 (2009).

15. Mueller, H. G., Weber, J. & Hornsby, B. W. Y. The Effects of Digital Noise Reduction on the Acceptance of Background Noise. *Trends in Amplification* **10**, 83–93 (2006).

16. Alcántara, J. L., Moore, B. C. J., Kühnel, V. & Launer, S. Evaluation of the noise reduction system in a commercial digital hearing aid. *International Journal of Audiology* **42**, 34–42 (2003).

17. Healy, E. W., Tan, K., Johnson, E. M. & Wang, D. An effectively causal deep learning algorithm to increase intelligibility in untrained noises for hearing-impaired listeners. *The Journal of the Acoustical Society of America* **149**, 3943–3953 (2021).

18. Goehring, T., Yang, X., Monaghan, J. J. M. & Bleeck, S. Speech enhancement for hearing-impaired listeners using deep neural networks with auditory-model based features. in *2016 24th European Signal Processing Conference (EUSIPCO)* 2300–2304 (2016).

19. Zhao, Y., Wang, D., Johnson, E. M. & Healy, E. W. A deep learning based segregation algorithm to increase speech intelligibility for hearing-impaired listeners in reverberant-noisy conditions. *The Journal of the Acoustical Society of America* **144**, 1627–1637 (2018).

20. Bramsløw, L. *et al.* Improving competing voices segregation for hearing impaired listeners using a low-latency deep neural network algorithm. *The Journal of the Acoustical Society of America* **144**, 172–185 (2018).

21. Healy, E. W. *et al.* Deep learning based speaker separation and dereverberation can generalize across different languages to improve intelligibility. *The Journal of the Acoustical Society of America* **150**, 2526–2538 (2021).

22. Goehring, T. *et al.* Speech enhancement based on neural networks improves speech intelligibility in noise for cochlear implant users. *Hearing Research* **344**, 183–194 (2017).

23. Andersen, A. H. *et al.* Creating Clarity in Noisy Environments by Using Deep Learning in Hearing Aids. *Semin Hear* **42**, 260–281 (2021).

24. Ronneberger, O., Fischer, P. & Brox, T. U-Net: Convolutional Networks for Biomedical Image Segmentation. *Medical Image Computing and Computer-Assisted Intervention – MICCAI 2015.* **9351**, (2015).

25. Elsken, T., Metzen, J. H. & Hutter, F. Neural Architecture Search: A Survey. *Journal of Machine Learning Research* **20**, 1–21 (2019).

26. Real, E., Aggarwal, A., Huang, Y. & Le, Q. V. Regularized Evolution for Image Classifier Architecture Search. *Proceedings of the AAAI Conference on Artificial Intelligence* **33**, (2019).

27. Howard, A. *et al.* Searching for MobileNetV3. *IEEE/CVF International Conference on Computer Vision (ICCV)* (2019).

28. Tan, M. *et al.* MnasNet: Platform-Aware Neural Architecture Search for Mobile. in *2019 IEEE/CVF Conference on Computer Vision and Pattern Recognition (CVPR)* 2815–2823 (2019).

29. Wang, Y. *et al.* Tacotron: Towards End-to-End Speech Synthesis. *INTERSPEECH 2017* (2017).

30. Oord, A. van den *et al.* WaveNet: A Generative Model for Raw Audio. *Proc. 9th ISCA Workshop on Speech Synthesis Workshop (SSW 9)* (2016).

31. Nicolson, A. & Paliwal, K. K. Masked multi-head self-attention for causal speech enhancement. *Speech Communication* **125**, 80–96 (2020).

32. Defossez, A., Synnaeve, G. & Adi, Y. Real Time Speech Enhancement in the Waveform Domain. *INTERSPEECH 2020* (2020).

33. Xu, R., Wu, R., Ishiwaka, Y., Vondrick, C. & Zheng, C. Listening to Sounds of Silence for Speech Denoising. *34th Conference on Neural Information Processing Systems (NeurIPS)* (2020).

34. Valentini-Botinhao, C. Noisy speech database for training speech enhancement algorithms and TTS models. *http://parole.loria.fr/DEMAND/* (2017).

35. Nagrani, A., Chung, J. S., Xie, W. & Zisserman, A. Voxceleb: Large-scale speaker verification in the wild. *Computer Speech & Language* **60**, 101027 (2020).

36. Wichern, G. *et al.* WHAM!: Extending Speech Separation to Noisy Environments. *INTERSPEECH 2019* (2019).

37. Kollmeier, B. & Wesselkamp, M. Development and evaluation of a German sentence test for objective and subjective speech intelligibility assessment. *The Journal of the Acoustical Society of America* **102**, 2412–2421 (1997).

38. Luo, Y. & Mesgarani, N. Conv-TasNet: Surpassing Ideal Time-Frequency Magnitude Masking for Speech Separation. *IEEE/ACM Transactions on Audio, Speech, and Language Processing* **27**, 1256–1266 (2019).

39. Luo, Y., Chen, Z. & Yoshioka, T. Dual-Path RNN: Efficient Long Sequence Modeling for Time-Domain Single-Channel Speech Separation. in *ICASSP 2020 - 2020 IEEE International Conference on Acoustics, Speech and Signal Processing (ICASSP)* 46–50 (2020).

40. Yu, D., Kolbæk, M., Tan, Z.-H. & Jensen, J. Permutation invariant training of deep models for speaker-independent multi-talker speech separation. in *2017 IEEE International Conference on Acoustics, Speech and Signal Processing (ICASSP)* 241–245 (2017).

41. Desjardins, J. L. The Effects of Hearing Aid Directional Microphone and Noise Reduction Processing on Listening Effort in Older Adults with Hearing Loss. *Journal of the American Academy of Audiology* **27**, 29–41 (2016).

42. Desjardins, J. L. & Doherty, K. A. The effect of hearing aid noise reduction on listening effort in hearing-impaired adults. *Ear and Hearing* **35**, 600–610 (2014).

43. Ohlenforst, B. *et al.* Impact of SNR, masker type and noise reduction processing on sentence recognition performance and listening effort as indicated by the pupil dilation response. *Hearing Research* **365**, 90–99 (2018).

44. Moore, G. E. Cramming more components onto integrated circuits. **38**, 4 (1965).

45. Hernandez, D. & Brown, T. Measuring the Algorithmic Efficiency of Neural Networks. *arXiv:2005.04305* (2020).

46. Almeida, M., Laskaridis, S., Leontiadis, I., Venieris, S. I. & Lane, N. D. EmBench: Quantifying Performance Variations of Deep Neural Networks across Modern Commodity Devices. *The 3rd International Workshop on Deep Learning for Mobile Systems and Applications - EMDL '19* 1–6 (2019).

47. He, K., Zhang, X., Ren, S. & Sun, J. Spatial Pyramid Pooling in Deep Convolutional Networks for Visual Recognition. *European Conference on Computer Vision (ECCV)* **8691**, 346–361 (2014).

48. Chollet, F. Xception: Deep Learning with Depthwise Separable Convolutions. in *2017 IEEE Conference on Computer Vision and Pattern Recognition (CVPR)* 1800–1807.

49. Loizou, P. C. *Speech Enhancement: Theory and Practice, Second Edition*. (CRC Press, 2013).

50. Taal, C. H., Hendriks, R. C., Heusdens, R. & Jensen, J. A short-time objective intelligibility measure for time-frequency weighted noisy speech. in *2010 IEEE International Conference on Acoustics, Speech and Signal Processing* 4214–4217 (2010).

51. Hines, A., Skoglund, J., Kokaram, A. C. & Harte, N. ViSQOL: an objective speech quality model. *EURASIP Journal on Audio, Speech, and Music Processing* **2015**, 13 (2015).

52. Rix, A. W., Beerends, J. G., Hollier, M. P. & Hekstra, A. P. Perceptual evaluation of speech quality (PESQ)-a new method for speech quality assessment of telephone networks and codecs. in *2001 IEEE International Conference on Acoustics, Speech, and Signal Processing. Proceedings* vol. 2 749–752 (2001).

53. Veaux, C., Yamagishi, J. & MacDonald, K. SUPERSEDED - CSTR VCTK Corpus: English Multi-speaker Corpus for CSTR Voice Cloning Toolkit. (2017).

54. Thiemann, J., Ito, N. & Vincent, E. DEMAND: a collection of multi-channel recordings of acoustic noise in diverse environments. (2013).

55. Wu, Y.-H. *et al.* Characteristics of Real-World Signal-to-noise Ratios and Speech Listening Situations of Older Adults with Mild-to-Moderate Hearing Loss. *Ear and hearing* **39**, 293–304 (2018).

56. scipy.signal.resample_poly — SciPy v1.6.3 Reference Guide. https://docs.scipy.org/doc/scipy/reference/generated/scipy.signal.resample_poly.html.

57. sox(1) - Linux man page. https://linux.die.net/man/1/sox.

58. facebookresearch/denoiser. https://github.com/facebookresearch/denoiser (2021).

59. anicolson/DeepXi. https://github.com/anicolson/DeepXi (2021).

60. Ephrat, A. *et al.* Looking to Listen at the Cocktail Party: A Speaker-Independent Audio-Visual Model for Speech Separation. *ACM Transactions on Graphics* **37**, 1–11 (2018).

61. Gemmeke, J. F. *et al.* Audio Set: An ontology and human-labelled dataset for audio events. in *2017 IEEE International Conference on Acoustics, Speech and Signal Processing (ICASSP)* 776–780 (2017).

62. Wagener, K. C. & Brand, T. Sentence intelligibility in noise for listeners with normal hearing and hearing impairment: Influence of measurement procedure and masking parameters. *International Journal of Audiology* **44**, 144–156 (2005).

# Methods

**Deep-learning based speech-enhancement**
The denoising system uses a deep neural network (also referred to as "network", Figure 1) with a U-Net[24] architecture. The U-Net consists of an encoder and a decoder separated by a bottleneck with skip connections running from encoder to decoder.

The encoder compresses the input using strided convolutions and the decoder reconstructs the compressed data back to its original dimensions using transposed convolutions. The encoder consists of 3 residual convolutional blocks, each containing several convolutional layers after which a strided convolution downsamples the data to decrease its dimensions. The blocks differ slightly in each stage and were selected by an evolutionary architecture search[25–28]. The search variables include the number of layers in each block, the type of layer, their kernel sizes, the number of filters, and dilation rate. One type of layer (e.g. 3x3 convolution) is repeated in each block and combined with a residual connection after each layer repetition. If necessary, an additional projection (dense) layer is used to adapt the feature dimensions of the skip connection to match the shape of the tensor it is being added to.

The decoder has its own 3 uniquely searched blocks, but with the same search space as the encoder. However, it replaces the down sampling strided convolutions after each block by a transposed convolution to up-sample the processed data back to its original dimensions. Additionally, the first residual connection of each block consists of a skip connection from the equivalent layer in the encoder.

The bottleneck consists of two parts: a convolutional block, as above but without strided or transposed convolutions, followed by a hand-designed pyramid pooling block[47].

The types of layers are limited to standard convolutional layers, depth-wise convolutional layers, spatial convolutions, and frequency/time first convolutions[48]. Spatially separable convolutions consist of subsequent distinct convolutional layers with variable kernel size in one spatial dimension (i.e. time or frequency), and a constant kernel size of 1 in the other dimensions. We distinguish two settings: 1) time first and 2) frequency first, to distinguish whether to apply a convolution in the time or frequency dimension first.

Throughout the network, each layer uses rectified linear unit (ReLu) activation functions except the last layer which uses a linear activation function. In total the network includes around 4 million parameters. We trained the network with a batch size of 16 samples and a length of 10 seconds (cut to 1.8 seconds to match the network input size) per sample, to predict the complex-valued ideal ratio mask of each sample using the mean squared error loss and optimize the weights with the Adam optimizer with a learning rate of 0.0001.

The network input consists of mixed speech and noise inputs taken from a large database of speech and noise files. Each mixed input waveform, sampled at 22.5kHz, is converted to the frequency domain using the short-time Fourier transform (STFT) which uses a frame length 512, frame step 128 and the Hann window function. The network predicts the denoised STFT which is then converted back to a waveform. During testing the waveform was sent to the soundcard of the testing device, which in this study was a laptop. The algorithm runs in real-time on Ubuntu 16.04 on a Asus FX504GD-DM116 laptop with the following specifications: an Intel Core i5-8300H, 8 GB DDR4 RAM, and a NVIDIA GTX 1050 graphics card. We did not attempt run-time optimizations of the model and instead focused on improving the network itself during development. It is to be expected that given effort in the implementation, similar network models should be able to run on significantly smaller systems.

**Deep-learning based speech-quality metric**
Our custom speech quality metric is generated by a multi-stage neural network, which was trained to predict human listeners' opinion scores from noisy speech files rated by human listeners on Amazon Mechanical Turk (MTurk).

**Ratings Dataset**
The MOS ratings were obtained using Amazon Mechanical Turk. To ensure that ratings were of high quality, only workers with a Mechanical Turk Master's qualification were accepted to work on the task. Workers rated sound files on a scale of 1.0 (bad) to 5.0 (excellent; see Extended Data Table 1) in batches of 17 samples. Among the 17 samples in each batch, we included 2 "baseline samples", one with a correct rating of 1.0 and the other with a correct rating of 5.0, which had to be rated correctly by the worker for the batch to be accepted into the dataset. As a note on the perceived quality when using a 1 to 5 scale MOS, the ceiling of performance for higher SNRs is caused by the limited possibility of improving the quality further (5 as highest rating) and the fact that even for the clean recorded samples the rating 5 is given only about 50% of the time.

The files to be rated were processed by a range of different denoising models. Models were selected to cover a large diversity in network architectures and training progress (i.e. models that were trained for only a few steps to models trained for millions of steps).

Altogether the final dataset consisted of 15,962 files, with 94,225 ratings, giving an average of 5.9 ratings per file. When calculating the average rating for each sample and category, the trimmed mean of all ratings is taken, i.e. the lowest and highest rating(s) were ignored.

**Synthetic ratings dataset**

To increase the size of the collected human-rated dataset, we trained a multilayer perceptron (MLP) to predict the MOS of each labelled sound file from a range of objective speech quality metrics, such as: SNR, mean squared error (MSE), mean log error, WSS, CEP (all described in[49]), STOI[50], ViSQOL[51], and PESQ[52]. We then used the trained MLP to predict the MOS from 600,000 unrated processed or unprocessed mixed speech and noise files from our datasets. We used this larger synthetic dataset of MLP-rated files to train our MOS prediction network described in the following section.

**Speech quality estimation network architecture**

The MOS network is an intrusive metric estimator receiving both a clean speech file and a corresponding (processed or unprocessed) mixed speech and noise file as inputs. The STFT and MEL spectrogram from the inputs is passed to the network, which is used to predict the MOS of the noisy sample. The network was trained on the synthetic dataset of 600,000 clean and noisy files described in the previous section.

Once trained, the MOS network evaluates the performance of the denoising network. The MOS network predicts one score for the unprocessed mixed speech and noise file and one for the same file after processing by the denoising network. The difference between the two scores in predicted MOS (delta-MOS) is taken as an evaluation metric. The evolutionary neural architecture search makes use of the delta-MOS as a target to find good hyperparameters and network structures for the denoising network.

**Denoising evaluation test sets**

We evaluated the performance of our denoising network and of several state-of-the-art models, which were either downloaded from the author's websites if pretrained models were available or reimplemented by ourselves following requests to the authors for available code.

We performed evaluations on 3 test sets: the Valentini[34] test set, commonly used to evaluate speech denoising models and 2 test sets consisting of publicly available VoxCeleb2[35] speech and WHAM! Noise[36] files, separated into WHAMVox easy, covering SNR ranges -12 to 27 dB, and WHAMVox hard, covering SNR ranges -20 to 20 dB. None of the files in the 3 test sets were included in the data used to train the models.

**Valentini test set**

The Valentini[34] dataset is available at https://datashare.ed.ac.uk/handle/10283/2791. The test set consists of 824 files containing speech samples from two speakers (one male and one female) taken from the VCTK[53] database and noises taken from the DEMAND[54] database. Each speech file is mixed with a noise sample at an SNR in the range of -1 dB to 17 dB.

**WHAMVox easy and hard test sets**

The WHAM![36] Noise set contains 3000 files (samples with pre-mixed speech and noise) from which we removed all samples above an estimated -12 dB SNR to reduce the probability of having audible speech in the noise files. Thus, the set was reduced to 1941 files with an average duration of ~10 seconds.

The VoxCeleb2[35] speech dataset contains ~36000 samples in its test set. To improve the quality of the files, we removed all samples with an estimated SNR below 20 dB and all files that are shorter than 8 seconds. This procedure reduced the number of audio files to ~5700 and left 116 distinct speakers in the test set, comprising 33 % female and 67% male speakers. All female speaker files and a randomly selected equal amount of male speaker files were selected to balance genders.

We controlled the SNR distribution of the mixed speech and noise files by sampling SNRs from a normal distribution with mean 8 dB SNR and standard deviation 7 that approximates the distribution of SNRs encountered in real world speech listening situations for older adults with mild to moderate hearing loss[55] (see Figure 4 of [55]). We then randomly sampled a SNR value, a noise file and a speaker file and mixed them together to give a total of 1941 mixed files with 69 distinct speakers (32 male, 37 female). We cropped files to their first 4 seconds to preserve the start of conversations and avoid audio files starting mid-sentence or word. For the WHAMVox hard test set we adjusted the SNR distribution, drawing from a normal distribution with mean 0 dB SNR and standard deviation 7. Before mixing, all files were normalized to have an RMS of –20 decibals relative to full scale (dBFS), controlling the loudness.

Code and setup requirements to create the WHAMVox test set are available online[‡].

All sound files were downsampled from their original sampling rate to 22.05kHz or 16kHz as expected by the denoising model. Downsampling was performed using scipy.signal.resample_poly[56] for all models except for the DEMUCS[32] model, where we used sox[57] to be consistent with the methods listed on the authors' website[58].

**Denoising comparison models**

We compared the results of our models to three state-of-the-art speech enhancement models. Pretrained DEMUCS[32] and MHANet[31] models were obtained from the authors' websites[58,59]. The third model we compared our model to, is the Sound of Silence[33] model. As no code was available at the time of writing and no pretrained model was available from the author's website, we reimplemented this model following the description from the paper[33].

---

[‡] The datasets can be downloaded from: https://audatic-team.github.io/WHAMVox/

Our implementation of the Sound of Silence model follows that described in the paper except for computation of the target silent interval mask, since in our experiments the original method performed poorly, mislabelling a high proportion of segments. Instead, we first took the STFT of the clean speech waveform, normalized the STFT magnitude to lie between -1 and 1 and applied the same threshold (0.08) to the STFT magnitude to produce the target mask. We trained the model end-to-end using an even weighting of all three target losses described in the original paper with the ADAM optimizer, learning rate set to 0.001, batch size 8, for 100,000 training steps. We trained the model using speech inputs drawn from the AVSPEECH dataset[60] and noise inputs drawn from the Audioset Noise (unbalanced) dataset[61]. All other model and training details are as specified in the original paper.

**Human evaluation of speech intelligibility**
The experimental protocols employed were approved by the Ethics Committee of the Charité Medical School Berlin, Germany and concur with the Helsinki Declaration. All subjects gave their informed consent.

**Hearing impaired study subjects**
Hearing impaired subjects were 16 German-speaking individuals (9 female, 7 male) aged 38-72 (mean 59.3 years, SD of +/- 8.3 years) with a binaural hearing loss of at least 30dB and varying degrees of experience with hearing aid use (0.8-20 years). A single subject had a middle-ear disorder in the left ear, while all other subjects had sensorineural hearing loss. The following exclusion criteria were applied: hearing aids or cochlear implants insufficiently fitted to understand speech in quiet; the usage of central nervous system drugs (e.g. anti-depressants, opioids) within 48 hours before the study participation.

**Normal hearing study subjects**
We tested 17 normal hearing German-speaking individuals (12 female, 5 male) aged 31 – 66 (mean 48.7 years, SD of +/- 11.3 years) without any hearing problem in the past. Usage of central nervous system drugs (e.g. anti-depressants, opioids) within 48 hours before the study participation was an exclusion criterion.

**Hearing test**
To determine the degree of hearing impairment, pure tone audiometry testing (0.25 to 8 kHz) was performed prior to the study. Hearing impaired subjects were tested under two conditions: once without hearing aid through air and bone conduction, and once with hearing aid, in the free field with the contralateral ear masked. In all conditions the pure-tone-average (PTA) was each determined using four test tones: 0.5, 1, 2 and 4 kHz.

**Subjects' hearing conditions**

Averaged over all patients, the air conduction hearing thresholds (Extended Data Figure 1) on the left ear are about 5 dB lower than on the right ear (except 1 kHz: 17 dB and 8 kHz: 12 dB). The median of the pure tone averages (PTA) of the unaided hearing is 50.6 dB on the right side and 42.5 dB on the left side. Statistical testing shows no significant difference between right and left side thresholds (p = 0.1; Wilcoxon test). In contrast, the hearing thresholds with hearing aids differ significantly between right and left side (p = 0.001) with lower thresholds on the left side (median of PTA - right: 32.5 dB, left: 25.6 dB).

According to the WHO grades of hearing loss[6] 1 had mild, 12 moderate, 1 moderately severe, 1 severe and 1 profound hearing loss.

The normal hearing study subjects all fulfill the WHO criteria for normal hearing[6] (the mean pure tone averages of 0.5, 1, 2 and 4 kHz are below 20 dB HL for both ears).

**OLSA Test**

We evaluated the effectiveness of our deep learning-based denoising algorithm using the Oldenburger Satztest[37] (OLSA test, Hörtech Inc., Germany), a German speech comprehension test in noise). The measured value is the speech reception threshold (SRT), which is the signal-to-noise ratio (SNR) at which 50% of the words in a sentence are understood correctly.

Mixed speech and noise were played through the loudspeaker at varying SNRs while the subject's SRT was measured. Speech consisted of OLSA sentences[37], with five words per sentence. These sentences are grammatically correct, without semantic cohesion. Each sentence was mixed with one of three types of noise: speech-shaped noise (original OLSA noise[62]), restaurant noise or traffic noise (both from our own datasets).

The noise level was fixed at 65 dB (SPL), while the speech level was adapted to determine the subject's SRT according to the OLSA procedure. At the beginning the speech level was also at 65 dB (SPL) and was then adjusted according to the number of words correctly repeated. If three or more words were repeated correctly, the speech level was decreased (according to a lookup-table by 1, 2 or 3 dB). If fewer than three words were repeated correctly, the level was increased. One test consisted of 20 sentences. The corresponding increases and decreases in speech level are such that the target word error rate is 50%.

In our study we performed with each subject 6 blocks of 20 sentences with different noise conditions: 1) OLSA noise[62]; 2) OLSA noise filtered by denoising system; 3) restaurant noise; 4) restaurant noise filtered by denoising system; 5) traffic noise; 6) traffic noise filtered by denoising system. The noises were presented in random order, but always starting with unfiltered noise followed directly by the sequence with the denoising system applied. The first 10 sentences of each block were ignored to allow for adaptation to the new test condition. The resulting SRT for each block was computed by taking the mean dB (SPL) of the final 10 sentences for that block.

The signal to noise ratio (SNR) was calculated by subtracting the speech level from the noise level (kept constant at 65 dB (SPL)).

The OLSA-Test requires each participant to be familiar with the procedure of the test. We therefore conducted a trial run with each subject before the measurements of the study.

During the measurements, each hearing impaired participant wore their own hearing aid, adjusted to their individual hearing profile.

**Test environment**
Subject testing took place in a soundproof room. The testing setup is shown in Extended Data Figure 2. A loudspeaker, one meter away from the subject, plays the speech and interference signals binaurally while a calibrated audiometer (SD 50, SIEMENS Inc., Germany) controls the sound levels of the speech and interference signals. OLSA test sentences were played from a given CD (Hörtech Inc., Germany), which contains balanced/randomized sentences. Our deep learning based denoising system modifies the mixed signal and is transferred through the soundcard (UM C22, Behringer Inc., Germany) to the loudspeaker. The devices were calibrated in accordance with international standards (EN 60645-1/-2/-4, EN ISO 389-1/-2/-3/-4/-5/-7, ISO 389-8) before the start of the study and the resulting sound level pressure was measured and kept constant before each study day.

**Statistical testing**
The distributions of the test results (SRT) are presented throughout as box-and-whiskers plots (outliers, 5-95% percentiles, 25-75% quartiles, median, mean as small triangle). For a comparison with literature data presented as mean and standard deviation the median absolute deviation (MAD) was calculated as a measure of variability (MAD = Median $\{|x_1 - M|, |x_2 - M|, |x_3 - M|, \ldots\}$ where M is the median of the series of data $\{x_1, x_2, x_3, \ldots\}$).

Statistical testing of SRT alterations between the two test conditions without or with the denoising system was performed with the Wilcoxon signed-rank test for matched pairs. To detect differences between several (more than 2) test conditions carried out with the same test group the Friedman test was used.

The comparison of test results between different test subjects was performed using the Mann-Whitney-U-test. In all tests a result of $p < 0.05$ was considered significant.

# Acknowledgements
The authors thank the Audatic team for their support. H.S. thanks the funding provided by the Deutsche Forschungsgemeinschaft (DFG, German Research Foundation) under Germany's Excellence Strategy – EXC 2002/1 "Science of Intelligence" – project number 390523135.

## Author contributions

P.U.D. designed the concept for this work. M.B., P.M.-R., and H.Z. created the denoising system. E.S. and P.U.D. supervised the creation of the denoising system. Y.S. and H.Z. implemented comparison denoising systems. P.M.-R., Y.S., and H.Z. conducted the online sound rating collection. P.U.D. supervised the online sound rating collection. V.M.H. and A.P. supervised the in-person studies with hearing impaired and normal hearing. U.S. and H.Z. conducted the in-person studies and performed the statistical analysis. P.U.D, Y.S., H.Z., U.S, designed the figures and illustrations. P.U.D, V.M.H., A.P., Y.S., H.Z., U.S, H.S., and E.S. contributed to the interpretation of the results. P.U.D, Y.S., H.Z., U.S, H.S., P.M.-R., E.S., A.P., V.M.H. wrote the manuscript. All authors provided critical feedback and helped revise the manuscript.